# Testing for the Presence of Structural Change and Spatial Heterogeneity

**Ruby Anne E. Lemence**

Bangko Sentral ng Pilipinas

**Erniel B. Barrios**

Professor, School of Statistics, University of the Philippines Diliman

**Abstract**   In a spatial-temporal model, structural change and/or spatial heterogeneity can easily affect estimation of parameters. Following the spatial-temporal model proposed by Landagan and Barrios (2007), we develop a nonparametric procedure for testing the presence of structural change and spatial heterogeneity using bootstrap techniques and the forward search algorithm. The time series bootstrap can filter the effect of temporary structural change in the construction of a confidence interval for the temporal parameter. The forward search will also facilitate the construction of a robust confidence interval for the spatial parameter. These confidence intervals are then used in deciding on the null hypothesis that there is no structural change/spatial heterogeneity. Simulation studies illustrate the ability of the proposed test procedure in detecting presence of structural change and spatial heterogeneity under certain conditions.

## 1. Introduction

Technological innovation led to tremendous improvements in data collection techniques allowing for data referenced both in space and in time. The information contained in such data are optimally utilized when models that take into account the dependence structure of data across space and over time are used, leading to spatial-temporal models.

Growing interests in modeling spatial-temporal systems in recent years is attributed to its emerging applications in the environmental and health sciences, which include monitoring of regional ozone levels, disease mapping, and analysis of satellite data, to name a few (Stroud et al., 2001). Spatial-temporal models are useful in monitoring and evaluation of policies, projects, interventions, etc., and the information they generate are useful tools in increasing public awareness as well as facilitating decision making (Landagan and Barrios, 2007). According to Stroud et al. (2001), another reason for the recent surge of interest in this area is the increased computational power as "space-time data sets are often large and therefore require substantial computing resources to fit even simple models."

It is not unlikely that we encounter data with atypical observations or sets of observations that seem to have a different structure compared with the rest. For time series data, unexpected events (shocks) may lead to changes in the parameters and these are exhibited by breaks in the series, for example, a shift in the mean of the series. Meanwhile, for data collected across space, some spatial units may have varying characteristics from the rest even if they are exposed to similar environmental settings.

In modeling, it is important that we take into account the observed changes in structure since failure to do so may lead to biased estimates of parameters. It is imperative to still estimate the process in the presence of such aberrant observations. Several procedures that enable robust estimation of the process even if there are atypical observations in the data have already been proposed, the forward search algorithm and bootstrap techniques included.

A method based on the forward search algorithm is proposed to detect spatial heterogeneity in spatial-temporal data using the model proposed by Landagan and Barrios (2007). A test for structural change based on bootstrap procedures, particularly the AR-sieve bootstrap for time series and the nonparametric bootstrap for independent observations is also proposed.

Deviations from the model assumptions usually occurs quite frequently whether in time or in space. However, most of the existing tests for structural change are based on asymptotic results. In a spatial-temporal setting, however, there is difficulty in coming up with large datasets that satisfy the requirements of the asymptotic procedures. In some cases, we may encounter data collected over a long period of time but only for a few spatial units or data collected for a short period of time but for a relatively large number of spatial units, thus, a need for tests that can address short time series data or small spatial samples. The bootstrap methods and the forward search algorithm are locally efficient methods, see for example (Beran, 1997) and (Atkinson and Riani, 2007); hence, these methods can work well in testing the presence of structural change and spatial heterogeneity in a typical spatial-temporal data.

We define structural change as the change in the value of the temporal parameter. A temporary structural change occurs if the value of the temporal parameter changed at some time points, then return to the previous values afterwards. On the other hand, spatial heterogeneity is present if for some spatial units, the value of the spatial parameter is different from the rest.

For structural change, we consider a situation where the temporal effect is different for some time periods compared with the rest, while assuming constant covariate effects across locations and time. Furthermore, we assume that the spatial effect is constant over time. Meanwhile, for spatial heterogeneity, we take into account a scenario where the spatial effect is different for some spatial units compared with the rest assuming no change in the covariate and temporal effects across space and time.

## 2. Testing for Structural Change and Spatial Heterogeneity

The spatial-temporal model proposed by Landagan and Barrios (2007) is given by the following:

$$Y_{it} = X_{it}\beta + W_{it}\delta + \varepsilon_{it}; i = 1,...,N; t = 1,...,T \tag{1}$$

where $Y_{it}$ is the response variable from location $i$ at time $t$, $X_{it}$ is a set of covariates from location $i$ at time $t$, $W_{it}$ is a set of variables in the neighborhood system from location $i$ at time $t$, and $\varepsilon_{it}$ is the error component. The error component is assumed to be autocorrelated (temporal dependence). Without loss of generality, assume that the error term follows an autoregressive process of order 1, given by, $\varepsilon_{it} = \rho\varepsilon_{it-1} + a_{it}$, $|\rho|<1$, $a_{it} \sim IID(0,\sigma_a^2)$. Further, model (1) assumes the following:

(i) constant covariate effect ($\beta$) across locations and time
(ii) constant temporal effect ($\rho$) across locations
(iii) constant spatial effect ($\delta$) across time

This study takes into account the following cases of deviations from the spatial-temporal model above: the temporal effect changed at specific time points; and the spatial effect is different for some groups of spatial units.

Modifying model (1), the spatial-temporal model with structural change and spatial heterogeneity is given by the following:

$$Y_{it} = X_{it}\beta + W_{it}\delta_h + \varepsilon_{it}, \; \varepsilon_{it} = \rho_c\varepsilon_{it-1} + a_{it}, \; |\rho|<1, a_{it} \sim iid(0,\sigma_a^2) \tag{2}$$

where $\delta_h = \delta I(i)_{\{i \notin N^h\}} + \delta' I(i)_{\{i \in N^h\}}$
$\rho_c = \rho I(i)_{\{i \notin T^c\}} + \rho' I(i)_{\{i \in T^c\}}$

$i=1, 2,...,N$ and $t=1,2,...,T$ and where δ and ρ are the original parameter values and δ′ and ρ′ are the new, temporary parameters values due to spatial heterogeneity and structural change, respectively. In addition, $N^h$ is the set of spatial units affected by heterogeneity, and $T^c$ is the set of time points where the temporary structural change occurred.

For each case, the proposed test methodology is composed of two phases: (i) estimation of model parameters and (ii) testing the null hypothesis of no structural change or no spatial heterogeneity. The AR-sieve bootstrap is used to replicate the process that generated the time series for each spatial unit. The purpose of the replication is to obtain as many estimates of the temporal parameter for subsequent use in the test. Even for independent observations, (Beran, 1997) proved the convergence of the intuitive bootstrap distribution to the true distribution function. On the other hand, (Bühlmann, 1997) argued that for as long as the AR(∞) model is correctly identified/estimated, then the AR-sieve is also consistent.

The forward search algorithm is used in obtaining robust estimates of the spatial parameters. The robustness of the forward search algorithm is discussed in (Atkinson and Riani, 2007) and further illustrated in (Atkinson, 2009). Robust estimates of spatial parameters are necessary since it is easily affected by spatial heterogeneity. The sampling distribution of the statistics to be tested will be determined through the bootstrap procedure. The empirical sampling distribution is then used to construct the 100(1-α)% confidence interval for the parameter of interest.

## 2.1. Testing for Structural Change

A structural change occurs if for some period, there is a considerable change in the values of the model parameters. The null hypothesis of no structural change will be tested against the alternative that a temporary structural change occurred at some unknown time point using bootstrap methods.

*Phase I. Estimation Algorithm using the AR-sieve Bootstrap*

For each spatial unit, $i$, $i = 1, ..., N$,

T.1. Estimate the parameters of the model, $Y_{it} = X_{it}\beta + W_{it}\delta + \varepsilon_{it}$ using ordinary least squares. Given the estimates, $\hat{\beta}$ and $\hat{\delta}$, compute the residuals $\{e_t\}$. The residuals $\{e_t\}$ contains information on the temporal dependencies, the autocorrelation structure indexed by ρ.

T.2. Fit an AR(1) model on $\{e_t\}$, the resulting parameter estimate is denoted by $\hat{\rho}$.

T.3. Given $\hat{\rho}$ obtained in (T.2), replicate $\{e_t\}$ with $\{e_t^*\}$ by generating error terms $\{a_t\}$ from $\hat{F}_{a,T}(.)$, the empirical distribution of the residuals in (T.2), $e_t^* = \hat{\rho} e_{t-1}^* + a_t$. Do this $m$ times to complete $m$ sets of $\{e_t^*\}$. Let $e = \{\{e_t^{*k}, t=1,..,T\}, k=1,...,m\}$.

T.4. For each of the $m$ simulated residuals in (T.3), generate pseudo-observations, $\{\{Y_t^{*1}\}, \{Y_t^{*2}\}, ..., \{Y_t^{*m}\}\}$, where

$$Y_t^{*k} = X_t \hat{\beta} + W_{it} \hat{\delta}_h + e_t^{*k},$$
$$k = 1, 2, ..., m.$$

T.5. For each of $\{Y_t^{*k}\}, k=1,2,...,m$, estimate the model $Y_t^{*k} = X_t \beta^{(k)} + W_{it} \delta_h^{(k)} + \varepsilon_t$, compute new residuals and fit an AR(1) model for the new residuals. Set aside the new parameter estimates.

### *Phase II. Bootstrap Test for Structural Change*

After exhausting all the $N$ spatial units, we now have *(N x m)* estimates for ρ. Ordinary bootstrap will then be performed using these *(N x m)* estimates to construct a confidence interval for $\rho$.

By construction, the AR(1) model in (2) depends heavily on the recent past. With the onset of structural change, the estimate of $\rho$ in T.2 will suffer. An estimate of $\rho$ that is unaffected by structural change is necessary so that the bootstrap confidence interval (BCI) to be used in the test is fairly robust and can detect such structural change more rationally. Otherwise, the structural change can push the BCI to widen, and will fail to recognize values of the estimates of $\rho$ that are different from the majority. The following bootstrap method will produce such BCI.

(i) From the *(N x m)* estimates for $\rho$ above, draw a simple random sample of size *n<Nxm* with replacement.

(ii)   For the sample obtained in *(i)*, compute the mean $\hat{\rho}^{(b)} = \frac{1}{n}\sum_{k=1}^{n}\hat{\rho}_k$ and the median $\hat{\rho}_{med}^{(b)}$.

(iii)  Repeat Steps *(i)* and *(ii)* B number of times, where B is a large number.

(iv)   Compute the bootstrap estimate for both the mean and the median using the formula, $\hat{\rho}_{BS} = \frac{1}{B}\sum_{j=1}^{B}\hat{\rho}^{(j)}$, Monte Carlo variance $\hat{\sigma}_{BS}^2 = \frac{1}{B-1}\sum_{j}^{B}(\hat{\rho}^{(j)} - \hat{\rho}_{BS})^2$, and 100(1-α)% BCI. The BCI is computed from the percentiles of the estimates of $\rho$ after (iii), see (Davison and Hinkley, 1997, pp. 202-211).

Conclude that there is no structural change if less than 100α% of the *(Nm)* $\hat{\rho}$ 's, fall outside the computed 100(1-α)% BCI for $\rho$. Otherwise, we reject the null hypothesis and conclude that a temporary structural change occurred.

The AR-sieve bootstrap and the classical bootstrap methods will ensure that the BCI for $\rho$ is fairly robust. The AR-sieve allows replication of the time series that will yield independent estimates of $\rho$ that are then subjected to classical bootstrap (assumes independent observations). The two levels of bootstrapping will ensure that initial effect of structural change will be neutralized until the BCI is constructed. When the estimates of $\rho$ for each spatial unit are compared to this interval, the presence of structural change will force some values outside the interval.

## *2.2. Testing for Spatial Heterogeneity*

When there is spatial heterogeneity, we will assume that the spatial parameter $\delta$ is different for some spatial units while assuming the covariate effects remain the same. The spatial unit manifesting the change is also assumed to be unknown.

The proposed procedure for testing spatial heterogeneity is also divided into two phases (estimation and detection. In the estimation phase, two types of estimates will be obtained: one, using the forward search algorithm and; another, using all the observations at once. The forward search algorithm will generate robust estimates of $\delta$ even when spatial heterogeneity is present. These "forward searched" estimates will then be bootstrapped to study the sampling distribution for $\hat{\delta}$. The estimates obtained using the full sample allows estimation of the actual process that generated the data.

The 100(1-α)% confidence interval for $\delta$ will be obtained using the empirical distribution (from bootstrap replicates) of $\hat{\delta}$. The estimates from the full sample will be compared with the confidence interval. Estimates outside the confidence interval provide evidence of spatial heterogeneity.

*Phase I. Estimation Using the Forward Search Algorithm*

The following steps will be performed for each time point, *t, t = 1,..., T*.
- S.1. From *N* observations, choose a subset of size *l*, *l<N*, such that it is outlier-free, thus, "ideal enough" to represent the *N* locations. This is done by fitting the model $Y_{it} = X_{it}\beta + W_{it}\delta_h + \varepsilon_{it}$ to the full dataset and choosing the *l* observations corresponding to the *l* smallest residuals.
- S.2. Fit the model in (S.1) to the selected *l* observations.
- S.3. Use the parameter estimates obtained in (S.2) to compute for the fitted $Y_{it}$ ($\hat{Y}_{it}$), for all $i = 1, 2, ..., N$ and obtain the residuals $e_{it} = Y_{it} - X_{it}\hat{\beta} - W_{it}\hat{\delta}_h$.
- S.4. Select the *(l+1)* observations that correspond to the *(l+1)* smallest residuals in (S.3).
- S.5. Fit the model in (S.2) to the *(l+1)* observations selected in (S.4).
- S.6. Repeat (S.2) to (S.5) adding one location at a time until all the *N* locations are included in the estimation.

In Atkinson and Riani (2007), the size of the initial subset was set to be equal to the number of parameters in the regression model considered. Meanwhile, in the simulation study discussed here, the size of the initial subset (*l*) was set to be equal to N/2 to minimize the fluctuations in Cook's *D* statistic used as the stopping criteria for the forward search algorithm. After fitting the model to the initial subset, the search progresses by moving on to a larger subset of observations, adding one observation at a time. In each iteration, the model in (S.1) is fitted to the subset and the estimates obtained are used to compute for the residuals using the full sample to determine the candidate observations to be included in the next iteration. We stop the search when the estimates are already "behaving wildly", i.e., when large changes are observed in the values of Cook's *D* statistic, meaning, newly included observation have undue influence on the regression coefficients, see for example, (Chatterjee, et. al., 2000, pp.104).

As the search progresses, the values of the Cook's *D* statistic are computed and observed for each iteration. The iteration is terminated if the change in the maximum value of the Cook's *D* statistic in the succeeding iteration exceeds some tolerance level (τ). The most recent estimate of the parameter is robust, and further inclusion of other "influential" observations will strongly influence subsequent estimates.

### *Phase II. Bootstrap Test for Spatial Heterogeneity*

From Phase I, we collect $T$ estimates of the spatial parameter $\delta$, given by $\hat{\underline{\delta}} = (\hat{\delta}_1, \hat{\delta}_2, ..., \hat{\delta}_T)$. To test the null hypothesis that there is no spatial heterogeneity, we follow similar bootstrap procedure as in testing for structural change. With spatial heterogeneity, some of the elements of $\hat{\underline{\delta}}$ would behave differently from the rest. This will be smoothened following this bootstrap procedure:

(i) From the $T$ estimates for $\delta$ obtained using forward search algorithm, draw a simple random sample of size $n<T$ with replacement.
(ii) For the sample in (i), compute the mean and median of $\hat{\underline{\delta}}$.
(iii) Repeat (i) and (ii) $B$ number of times, where $B$ is a large number.
(iv) Compute the bootstrap estimate for both the mean and the median using the formula, $\hat{\delta}_{BS} = \frac{1}{B}\sum_{j=1}^{B} \hat{\delta}^{(j)}$, Monte Carlo variance $\hat{\sigma}^2_{BS} = \frac{1}{B-1}\sum_{j}^{B}(\hat{\delta}^{(j)} - \hat{\delta}_{BS})^2$, and 100(1-α)% bootstrap confidence interval (BCI) for $\delta$ from the percentiles generated in (iii).

Conclude that there is spatial heterogeneity if more than 100α% of the $T$ $\hat{\delta}$'s, computed using the full sample, fall outside the 100(1-α)% BCI for $\delta$. Otherwise, do not reject the null hypothesis and conclude that there is no spatial heterogeneity. Since the forward searched and bootstrapped estimates of $\delta$ are robust to spatial heterogeneity, the resulting bootstrap confidence interval would make a reasonable benchmark in deciding which hypothesis to favor that best reflects the nature of the data.

## 2.3. Simultaneous Testing for Structural Change and Spatial Heterogeneity

When both structural change and spatial heterogeneity occurs, the methods presented in Sections 4.2.1 and 4.2.2 need to be implemented iteratively. In Section 4.2.1, when spatial heterogeneity is ignored, the AR-sieve estimates will be affected. Similarly, when structural change is ignored, the forward searched estimates will be affected. Following the backfitting iterations in an additive model in (2), Phase 1 in Sections 4.2.1 and 4.2.2 will produce 'optimal' estimates. (Buja, et. al., 1989) proved consistency and convergence of the backfitting algorithm in a relatively general class of smoothers in an additive model. Thus, in Sections 4.2.1 and 4.2.2, only Phase 1 is to be implemented while the iteration is on-going. This will ensure the generation of robust estimates of the model parameters. Upon convergence, Phase 2 in both sections should be able to test the simultaneous occurrences of structural change and spatial heterogeneity.

## 3. Simulation Study

A simulation study is designed to evaluate the proposed test. Data that does not contain structural change/spatial heterogeneity are generated to assess if the tests are correctly sized. On the other hand, data that actually contained structural change and/or spatial heterogeneity are used in evaluating power of the test, i.e., whether the test can correctly detect these data aberrations.

The response variable *Y* was computed using the following model:
$$Y_{it} = \beta_0 + \beta_1 X_{1it} + \beta_2 X_{2it} + \delta_h W_{it} + \varepsilon_{it}$$
$$\varepsilon_{it} = \rho \varepsilon_{it-1} + a_{it} \text{ where } a_{it} \sim N(0,4)$$

where $X_1$ and $X_2$ were each sampled from the Normal population with means 100 and 50, respectively, and variances which are both equal to 100. To induce spatial dependencies, the total number of spatial units was divided into four clusters (neighborhood) through sampling of the spatial variable *W* from Poisson distribution with means $\lambda_k$, $\lambda_k \in \{2, 4, 6, 10\}$. Each $\lambda_k$ corresponds to one neighborhood or cluster. To illustrate, suppose spatial units are farmers whose yield (*Y*) are observed quarterly (*t*) along with $X_1$ number of years farming, $X_2$ area cultivated, and *W*, the number of times rainfall amount exceeds a threshold leading to a weeklong flooding in the farm. The sample farmers were gathered from four different regions, with varying rainfall patterns. Thus, one group had average of 2 weeklong flooding, another group with average of 4, another with 6, and the other with 10 weeklong flooding per quarter. The number of weeklong flooding will have proportional effect on the yield (constant δ), but could also pull the yield down tremendously (first three groups with constant δ, the fourth with δ′) when weeklong flooding happened very frequently, thus, the spatial heterogeneity. Meanwhile, the error term was simulated from the AR(1) process, $\varepsilon_{it} = \rho * \varepsilon_{it-1} + a_{it}$, with $\rho$ equal to 0.5, and $a_{it}$ was sampled from the Normal distribution with zero mean and variance equal to 4. The values of the coefficients used are as follows: $\beta_0 = 40.00; \beta_1 = 0.70; \beta_2 = 0.45; and, \delta = 0.25$. To induce structural change and spatial heterogeneity in the space-time data, $\rho = 0.75$ and $\delta = 1.25$ were considered as the temporary values. The error terms $a_{it}$ are the same whether or not a structural change happened.

The choice of the population where in the covariates and spatial variables are drawn is arbitrary and will not affect the evaluation of the procedure. In addition, we considered simulating datasets for different values of the coefficient of determination ($R^2$) to assess the effect of model fit. The values for $R^2$ considered were in the vicinity of 20%, 50%, and 95%, to represent low, moderately high, and high $R^2$.

### *3.1. Simulating Structural Change*

Consider a space-time dataset given by $\underline{Y} = \{Y_{it} : i = 1, 2, ..., N, t = 1, 2, ..., T\}$. In matrix notation,

$$\underline{Y} = \begin{pmatrix} y_{11} & y_{21} & \cdots & y_{N1} \\ y_{12} & y_{22} & \cdots & y_{N2} \\ \vdots & \vdots & \ddots & \vdots \\ y_{1T} & y_{2T} & \cdots & y_{NT} \end{pmatrix}$$

and note that $\underline{Y}$ is composed of N (Tx1) vectors,

$$\underset{\sim}{y}_i = \begin{bmatrix} y_{i1} \\ y_{i2} \\ \vdots \\ y_{iT} \end{bmatrix}, \ i = 1, 2, ..., N.$$

Three cases of structural change were considered depending on which part of the time series it occurred: start, middle, or end. Furthermore, it is assumed that if a temporary structural change occurs, all of the N spatial units are affected and that the occurrence of structural change is the same across space (all neighborhoods).

Structural change was introduced into the data by simply replacing the parameter of the autocorrelated errors for the selected period of the series keeping the same order of the autoregressive process, in this case, from ρ=0.5, it is changed to ρ=0.75 for portions of the time series where a structural change occurred supposedly. Furthermore, since the structural change is temporary, only 5%, 10%, and 15% of the given number of time points were allowed to be affected. Also, to determine the effect of sample size, we considered 40, 50, and 75 time points to represent small, medium and large sample sizes. The number of spatial units was also varied.

Meanwhile, for the AR-sieve bootstrap, 100 replicates were obtained for each spatial unit, *i*. 1000 resamples with replacement were considered for the ordinary bootstrap used in the construction of the 100(1-α)% bootstrap confidence interval.

## 3.2. Simulating Spatial Heterogeneity

Consider the space-time data set given by $\underline{Y} = \{Y_{it} : i = 1, 2, ..., N, t = 1, 2, ..., T\}$, whose matrix notation is also given in Section 6.1. $\underline{Y}$ is composed of *T (Nx1)* vectors, $\underline{y}_t = \begin{bmatrix} y_{1t} \\ y_{2t} \\ \vdots \\ y_{Nt} \end{bmatrix}$, $t = 1, 2, ..., T$.

Spatial dependence was introduced by dividing the data collected across space into four neighborhoods given by $\underline{Y} = (\underline{Y}1, \underline{Y}2, \underline{Y}3, \underline{Y}4)$ where *Y1* is associated with the neighborhood where *W~Po(2)*, *Y2* is associated with the neighborhood where *W~Po(4)*, *Y3* is associated with the neighborhood where *W~Po(6)*, and *Y4* is associated with the neighborhood where *W~Po(10)*. A neighborhood is defined as a group of spatial units that shares some common characteristics in the spatial variable *W*.

Four scenarios of spatial heterogeneity were considered depending on the number of neighborhoods affected. The number of affected spatial units was distributed in one, two, three, or all four neighborhoods. The proportions of spatial units exhibiting spatial heterogeneity considered were 5%, 10%, and 15%.

Spatial heterogeneity was induced for each set of spatial units, collected over time, by changing the value of the spatial parameter in the selected spatial units from 0.25 to 1.25. The effect of sample size was investigated by considering 20, 40, and 60 as the number of spatial units available. In addition, the number of time points was also allowed to vary. In the conduct of the ordinary bootstrap, 1000 bootstrap samples of size *T* were considered.

## 4. Results and Discussion

### *4.1. Test for Structural Change*

The test is correctly sized, since it accepts the null hypothesis of no structural change when in fact, there is no structural change. The method is also powerful in detecting structural change since it rejects the null hypothesis of no temporary structural change if indeed the data is embedded with structural change.

For structural change that occurred at the start of the series, the test appears to be powerful whether the sample size is small or large. On the other hand, for structural change that occurred in the middle, the test is powerful provided that the sample size is small to moderate and the proportion of observations with structural change is relatively large. Meanwhile, for structural change that occurred at the end of the series, the test is powerful in detecting structural change provided that the proportion of observations exhibiting the change is not that large. See Table 1 for details.

Autoregressive models put more weight to more recent observations than older ones, thus, the procedure can easily detect temporary structural change that occur at the start of the series because the structure of the observations at the start is different from the more recent observations. Meanwhile, the procedure can only detect temporary structural change that occurs at the end of the series if the proportion of observations with structural change is low to moderate since if the extent of structural change is large, then, given that the AR estimation puts more weight to recent observations, it considers the structure of the new observations as the "true" pattern of the data and not just a temporary structural change. The extent of temporary structural change should be large enough for the procedure to detect it when it occurred at the middle of the series because the weights on the observations at the middle of the series can either be small or large depending on the length of the time series.

The test finds it difficult to correctly decide that there is no temporary structural change when there is really none when the number of spatial units is small and the number of time points is small to moderately large. Similarly, the abovementioned difficulty is observed as the number of spatial units is increased and the number of times points available is large.

**Table 1. Summary of the Results for the Test for Temporary Structural Change: $R^2=95\%$**

(Percentage of $\hat{\rho}'s$ covered in the 95% Bootstrap Confidence Interval for $\rho$)

| Position of Structural Change | N=20 | | | N=40 | | | N=60 | | |
|---|---|---|---|---|---|---|---|---|---|
| | T=40 | T=50 | T=75 | T=40 | T=50 | T=75 | T=40 | T=50 | T=75 |
| No Structural Change | 92.2 | 95.0 | 95.5 | 95.0 | 96.9 | 94.2 | 95.9 | 95.4 | 93.1 |
| (a) Proportion of time points with structural change = 5% | | | | | | | | | |
| Start | 94.3 | 93.5 | 94.6 | 94.2 | 93.1 | 93.7 | 95.2 | 96.4 | 95.0 |
| Middle | 95.0 | 97.4 | 94.2 | 95.9 | 95.7 | 95.3 | 95.3 | 95.7 | 95.0 |
| End | 93.9 | 94.6 | 93.7 | 95.3 | 94.8 | 93.8 | 95.4 | 94.4 | 95.5 |
| (b) Proportion of time points with structural change = 10% | | | | | | | | | |
| Start | 92.8 | 96.4 | 94.7 | 95.0 | 95.1 | 93.4 | 95.7 | 95.5 | 95.8 |
| Middle | 95.6 | 94.5 | 95.7 | 94.6 | 95.2 | 96.3 | 95.2 | 95.9 | 94.8 |
| End | 93.1 | 92.6 | 94.3 | 94.8 | 94.1 | 95.2 | 96.4 | 94.7 | 95.0 |
| (c) Proportion of time points with structural change = 15% | | | | | | | | | |
| Start | 94.3 | 94.2 | 94.8 | 92.6 | 95.0 | 94.8 | 94.9 | 95.3 | 95.8 |
| Middle | 93.7 | 94.9 | 95.4 | 96.2 | 94.5 | 95.2 | 95.6 | 94.7 | 93.5 |
| End | 96.5 | 95.9 | 96.0 | 95.6 | 95.5 | 94.1 | 95.0 | 95.7 | 95.5 |

Notes:

1. $H_o$: There is no temporary structural change vs. $H_a$: A temporary structural change occurred. The null hypothesis is rejected when the percentage of $\hat{\rho}'s$ covered in the 95% bootstrap confidence interval is less than 95%.

2. When there is no structural change, the test is correctly sized if the null hypothesis is not rejected. On the other hand, if a structural change occurred at the start, middle or end of the series, the test is powerful if the null hypothesis is rejected.

Meanwhile, when the number of spatial units is large and there is relatively fewer number of time points, the test can only detect the temporary structural change if it occurs at the start of the series and the proportion of affected observations is large. Temporary structural change that occurs in the middle or at the end of the series is easily detected as the number of time points is increased for large number of spatial units. It was noted earlier that temporary structural change that occurs in the middle of the series can easily be detected as long as the proportion of affected time points is large; however, there is also evidence that when both the number of spatial units and time points are large, the temporary structural change that occurs in the middle is easily detected regardless of the size of the affected time points. When the number of spatial units and the number of time points are both large, the temporary structural change occurring at the start or at the end of the series become negligible and can no longer be detected by the test.

The performance of the proposed procedure was also evaluated for cases when the fit of the model is good/poor by considering high/low values of the coefficient of determination, $R^2$ gauged from the data-generating mechanism. The values of $R^2$ considered were in the vicinity of 20% and 95%. The procedure is unaffected by the change in the level of $R^2$. See Table 2 for details. Multiplying a constant to the error term resulted in higher variance but the autoregressive parameter remained the same. Thus, it can be observed from the tables that the median of the estimates, as well as the low- and high-end values of the 95% BCI for the temporal parameter are similar for both levels of $R^2$.

**Table 2. Summary of the Results of the Test for Temporary Structural Change $R^2$=20%**

(Percentage of $\hat{\rho}'s$ covered in the 95% Bootstrap Confidence Interval for $\rho$)

| Position of Structural Change | N=20 | | | N=40 | | | N=60 | | |
|---|---|---|---|---|---|---|---|---|---|
| | T=40 | T=50 | T=75 | T=40 | T=50 | T=75 | T=40 | T=50 | T=75 |
| No Structural Change | 92.2 | 94.5 | 94.4 | 95.0 | 96.9 | 94.2 | 95.9 | 95.4 | 93.1 |
| (a) Proportion of time points with structural change = 5% | | | | | | | | | |
| Start | 94.3 | 93.5 | 95.0 | 93.8 | 93.1 | 93.7 | 95.2 | 96.4 | 95.0 |
| Middle | 95.0 | 97.1 | 93.8 | 95.9 | 95.7 | 95.3 | 95.3 | 95.7 | 95.0 |
| End | 94.0 | 94.7 | 94.0 | 95.7 | 94.8 | 93.8 | 95.4 | 94.4 | 95.5 |
| (b) Proportion of time points with structural change = 10% | | | | | | | | | |
| Start | 92.8 | 96.1 | 95.1 | 95.0 | 95.1 | 93.4 | 95.7 | 95.5 | 95.8 |
| Middle | 95.6 | 93.9 | 96.3 | 94.6 | 95.2 | 96.3 | 95.2 | 95.9 | 94.8 |
| End | 93.0 | 92.8 | 94.3 | 94.0 | 94.1 | 95.1 | 96.4 | 94.7 | 95.0 |
| (c) Proportion of time points with structural change = 15% | | | | | | | | | |
| Start | 94.3 | 94.2 | 93.8 | 92.6 | 95.0 | 94.8 | 94.9 | 95.3 | 95.8 |
| Middle | 93.7 | 96.0 | 95.4 | 96.2 | 94.5 | 95.2 | 95.6 | 94.7 | 93.5 |
| End | 97.3 | 96.0 | 95.3 | 96.5 | 95.5 | 94.1 | 95.0 | 95.7 | 95.5 |

## 4.2. Test for Spatial Heterogeneity

The test for spatial heterogeneity is also correctly sized since it accepts the null hypothesis of no spatial heterogeneity when there is none. It is also powerful since it rejects the null hypothesis of no spatial heterogeneity when the data is embedded with spatial heterogeneity.

When the fit of the model is good and spatial heterogeneity is actually present, it is easily detected when the number of spatial units is small and the number of time points is relatively large. The number of neighborhoods affected with spatial heterogeneity does not matter. On the contrary, for large number of spatial units, for small to moderately large number of time points, spatial heterogeneity must be present in all of the neighborhoods in order for the test to detect it. Meanwhile, as the number of spatial units and number of time points become large, the procedure can no longer detect spatial heterogeneity as the change in the spatial effect is diluted with more observations with no spatial heterogeneity. See Table 3 for details.

**Table 3 Summary of the Results of the Tests for Spatial Heterogeneity $R^2$=95%.**

(Percentage of $\hat{\delta}'s$ covered in the 95% Bootstrap Confidence Interval for $\delta$ )

| No. of Neighborhoods with Spatial Heterogeneity | N=20 | | | N=40 | | | N=60 | | |
|---|---|---|---|---|---|---|---|---|---|
| | T=40 | T=50 | T=75 | T=40 | T=50 | T=75 | T=40 | T=50 | T=75 |
| No Spatial Heterogeneity | 95.0 | 94.0 | 96.0 | 97.5 | 96.0 | 97.3 | 92.5 | 96.0 | 100.0 |
| (a) Proportion of spatial units with spatial heterogeneity = 5% | | | | | | | | | |
| 1 Neighborhood | 92.5 | 88.0 | 97.3 | 92.5 | 98.0 | 98.7 | 100.0 | 100.0 | 98.7 |
| 2 Neighborhoods | 95.0 | 100.0 | 96.0 | 92.5 | 94.0 | 94.7 | 100.0 | 100.0 | 100.0 |
| 3 Neighborhoods | 92.5 | 96.0 | 97.3 | 95.0 | 94.0 | 93.3 | 97.5 | 92.0 | 98.7 |
| 4 Neighborhoods | 90.0 | 88.0 | 94.7 | 95.0 | 92.0 | 93.3 | 92.5 | 94.0 | 96.0 |
| (b) Proportion of spatial units with spatial heterogeneity = 10% | | | | | | | | | |
| 1 Neighborhood | 95.0 | 94.0 | 97.3 | 95.0 | 98.0 | 96.0 | 100.0 | 94.0 | 97.3 |
| 2 Neighborhoods | 95.0 | 94.0 | 94.7 | 92.5 | 96.0 | 94.7 | 97.5 | 98.0 | 97.3 |
| 3 Neighborhoods | 100.0 | 98.0 | 94.7 | 95.0 | 96.0 | 92.0 | 100.0 | 100.0 | 100.0 |
| 4 Neighborhoods | 92.5 | 94.0 | 93.3 | 90.0 | 90.0 | 90.7 | 92.5 | 94.0 | 94.7 |
| (c) Proportion of spatial units with spatial heterogeneity = 15% | | | | | | | | | |
| 1 Neighborhood | 90.0 | 98.0 | 89.3 | 100.0 | 94.0 | 94.7 | 95.0 | 96.0 | 97.3 |
| 2 Neighborhoods | 97.5 | 96.0 | 90.7 | 92.5 | 98.0 | 96.0 | 100.0 | 98.0 | 98.7 |
| 3 Neighborhoods | 95.0 | 96.0 | 93.3 | 90.0 | 90.0 | 92.0 | 97.5 | 92.0 | 98.7 |
| 4 Neighborhoods | 100.0 | 100.0 | 96.0 | 87.5 | 84.0 | 93.3 | 92.5 | 94.0 | 100.0 |

When the number of time points is small to moderately large and the number of spatial units is small, the proportion of units with spatial heterogeneity must be small to moderately large for the procedure to detect spatial heterogeneity. However, for large number of time points, the extent of spatial heterogeneity must be large enough for it to be detected. For moderately large number of spatial units (N=40), it seems that the test is sensitive for the presence of spatial heterogeneity regardless of the number of times points available. For large number of spatial units, the number of time points should not be very large and spatial heter-ogeneity must be present in almost all neighborhoods for the procedure to detect it.

When the fit of the model is poor, the test is still correctly sized (see Tables 4 and 5). However, for $R^2=50\%$, it was observed that the procedure already finds it difficult to detect spatial heterogeneity when the number of spatial units is small. When the number of spatial units is small, spatial heterogeneity is detected only when it is present in all of the neighborhoods and the number of time points is large enough. The procedure is powerful in detecting spatial heterogeneity for moderately large number of spatial units (N=40), provided that the extent of spatial heterogeneity is large enough. Meanwhile, for large number of spatial units, spatial heterogeneity is detected when it is scattered across the neighborhoods and the extent of spatial heterogeneity is large enough. Still for $R^2 = 20\%$, the procedure is sensitive to spatial heterogeneity when the extent of spatial heterogeneity is at most moderately large for small $N$ and moderately large to large $T$ and for small $T$ when $N$ is relatively large. In addition, when the fit of the model is not good, there is a possibility that the spatial effect is not significantly different from zero and the 95% BCI's obtained are very wide.

**Table 4 Summary of the Results of the Test for Spatial Heterogeneity: $R^2=50\%$.**

(Percentage of $\hat{\delta}'s$ covered in the 95% Bootstrap Confidence Interval for $\delta$)

| No. of Neighborhoods with Spatial Heterogeneity | N=20 | | | N=40 | | | N=60 | | |
|---|---|---|---|---|---|---|---|---|---|
| | T=40 | T=50 | T=75 | T=40 | T=50 | T=75 | T=40 | T=50 | T=75 |
| No Spatial Heterogeneity | 95.0 | 94.0 | 97.3 | 97.5 | 96.0 | 94.7 | 92.5 | 96.0 | 100.0 |
| (a) Proportion of spatial units with spatial heterogeneity = 5% | | | | | | | | | |
| 1 Neighborhood | 97.5 | 98.0 | 96.0 | 92.5 | 98.0 | 97.3 | 95.0 | 96.0 | 97.3 |
| 2 Neighborhoods | 95.0 | 96.0 | 96.0 | 92.5 | 92.0 | 97.3 | 92.5 | 100.0 | 93.3 |
| 3 Neighborhoods | 95.0 | 96.0 | 97.3 | 92.5 | 96.0 | 96.0 | 100.0 | 100.0 | 100.0 |
| 4 Neighborhoods | 95.0 | 94.0 | 97.3 | 97.5 | 96.0 | 96.0 | 95.0 | 98.0 | 98.7 |
| (b) Proportion of spatial units with spatial heterogeneity = 10% | | | | | | | | | |
| 1 Neighborhood | 97.5 | 100.0 | 97.3 | 92.5 | 92.0 | 93.3 | 100.0 | 100.0 | 97.3 |
| 2 Neighborhoods | 97.5 | 96.0 | 97.3 | 92.5 | 94.0 | 98.7 | 92.5 | 100.0 | 96.0 |
| 3 Neighborhoods | 95.0 | 98.0 | 96.0 | 92.5 | 96.0 | 97.3 | 92.5 | 100.0 | 96.0 |
| 4 Neighborhoods | 95.0 | 94.0 | 97.3 | 97.5 | 96.0 | 94.7 | 95.0 | 92.0 | 94.7 |
| (a) Proportion of spatial units with spatial heterogeneity = 15% | | | | | | | | | |
| 1 Neighborhood | 100.0 | 98.0 | 94.7 | 92.5 | 96.0 | 97.3 | 97.5 | 100.0 | 100.0 |
| 2 Neighborhoods | 97.5 | 96.0 | 97.3 | 97.5 | 98.0 | 98.7 | 100.0 | 100.0 | 100.0 |
| 3 Neighborhoods | 97.5 | 96.0 | 96.0 | 92.5 | 96.0 | 98.7 | 95.0 | 96.0 | 94.7 |
| 4 Neighborhoods | 95.0 | 96.0 | 97.3 | 92.5 | 98.0 | 96.0 | 95.0 | 94.0 | 90.7 |

**Table 5 Summary of the Results of the Tests for Spatial Heterogeneity: R²=20%.**

(Percentage of $\hat{\delta}'s$ covered in the 95% Bootstrap Confidence Interval for $\delta$)

| No. of Neighborhoods with Spatial Heterogeneity | N=20 | | | N=40 | | | N=60 | | |
|---|---|---|---|---|---|---|---|---|---|
| | T=40 | T=50 | T=75 | T=40 | T=50 | T=75 | T=40 | T=50 | T=75 |
| No Spatial Heterogeneity | 95.0 | 94.0 | 97.3 | 97.5 | 96.0 | 94.7 | 92.5 | 96.0 | 100.0 |
| (a) Proportion of spatial units with spatial heterogeneity = 5% | | | | | | | | | |
| 1 Neighborhood | 95.0 | 96.0 | 92.0 | 92.5 | 98.0 | 97.3 | 92.5 | 92.0 | 98.7 |
| 2 Neighborhoods | 95.0 | 94.0 | 96.0 | 92.5 | 96.0 | 97.3 | 92.5 | 94.0 | 96.0 |
| 3 Neighborhoods | 95.0 | 94.0 | 97.3 | 92.5 | 96.0 | 93.3 | 97.5 | 100.0 | 98.7 |
| 4 Neighborhoods | 95.0 | 96.0 | 92.0 | 95.0 | 96.0 | 96.0 | 97.5 | 98.0 | 100.0 |
| (b) Proportion of spatial units with spatial heterogeneity = 10% | | | | | | | | | |
| 1 Neighborhood | 97.5 | 100.0 | 96.0 | 90.0 | 96.0 | 98.7 | 95.0 | 98.0 | 100.0 |
| 2 Neighborhoods | 95.0 | 96.0 | 97.3 | 90.0 | 98.0 | 97.3 | 92.5 | 100.0 | 100.0 |
| 3 Neighborhoods | 95.0 | 94.0 | 97.3 | 97.5 | 96.0 | 96.0 | 95.0 | 100.0 | 100.0 |
| 4 Neighborhoods | 95.0 | 96.0 | 92.0 | 97.5 | 96.0 | 97.3 | 97.5 | 100.0 | 94.7 |
| (a) Proportion of spatial units with spatial heterogeneity = 15% | | | | | | | | | |
| 1 Neighborhoods | 97.5 | 98.0 | 96.0 | 95.0 | 98.0 | 96.0 | 97.5 | 100.0 | 100.0 |
| 2 Neighborhoods | 97.5 | 96.0 | 98.7 | 92.5 | 98.0 | 97.3 | 95.0 | 100.0 | 100.0 |
| 3 Neighborhoods | 95.0 | 96.0 | 97.3 | 95.0 | 96.0 | 96.0 | 100.0 | 100.0 | 100.0 |
| 4 Neighborhoods | 95.0 | 96.0 | 97.3 | 95.0 | 96.0 | 96.0 | 97.5 | 96.0 | 94.7 |

# 5. Conclusions

A procedure based on the AR-sieve bootstrap was used to detect temporary structural change and a procedure based on the forward search algorithm was used to detect spatial heterogeneity is a spatial-temporal model. Bootstrap procedures were used to construct the confidence intervals for the temporal and spatial parameters, which, in turn, were used to test whether a temporary structural change or spatial heterogeneity is present or not. The tests are both correctly sized and powerful for a wide range of scenarios. The nonparametric procedure for testing temporary structural change is recommended when one suspect that there has been a change in the temporal effect with recent observations and the available space-time data is composed of a relatively short time series and relatively small number of spatial units. On the other hand, the use of the nonparametric procedure for testing spatial heterogeneity is recommended when dealing with relatively small to moderate number of spatial units and assuming that the fit of model is good enough.